\documentclass[a4paper,11pt]{article}
\usepackage{pos}
\usepackage{etoolbox}
\usepackage{setspace}
\makeatletter
\patchcmd{\thebibliography}
  {\settowidth}
  {\setlength{\itemsep}{0pt}%
   \setlength{\parskip}{0pt}%
   \settowidth}{}{}
\makeatother

\def\Journal#1#2#3#4{{#1} {\bf #2}, #3 (#4)}

\title{Radio Signatures of Cosmic-Ray Particle Showers in Deep In-Ice Antennas}
\author*[a]{Simon Chiche}
\author[b]{Krijn D. de Vries}
\author[a]{Simona Toscano}

\affiliation[a]{Inter-University Institute For High Energies (IIHE), Université libre de Bruxelles (ULB), \\
Boulevard du Triomphe 2, 1050 Brussels, Belgium}

\affiliation[b]{Vrije Universiteit Brussel (VUB), Dienst ELEM, Pleinlaan 2, B-1050, Brussels, Belgium}

\emailAdd{simon.chiche@ulb.be}

\abstract{To detect ultra-high-energy neutrinos, experiments such as ARA and RNO-G target the radio emission induced by these particles as they cascade in the ice, using deep in-ice antennas at the South Pole or in Greenland. In this context, it is essential to first characterize the in-ice radio signature from cosmic-ray-induced particle showers, which constitute a primary background for neutrino detection, and represent the fist in-situ detection of  in-ice particle cascades with radio antennas. This characterization will help validate the detection principle and assist in calibration.  To achieve this goal, we used FAERIE, the ``Framework for the simulation of Air shower Emission of Radio for in-Ice Experiments'', that combines CoREAS and GEANT4 to simulate the radio emission of cosmic ray showers deep in the ice. Using this tool, we analyze in-ice radio signatures of cosmic-ray showers, including polarization, timing, and radiation energy, as well as their dependence on shower parameters. These insights will facilitate the first cosmic-ray detections and improve cosmic-ray/neutrino discrimination.}

\ConferenceLogo{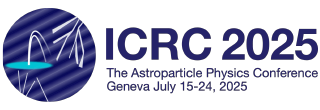}

\FullConference{39th International Cosmic Ray Conference (ICRC2025)\\
 15–24 July 2025\\
Geneva, Switzerland\\}
\begin{document}
\maketitle

\section{Introduction}
\vspace{-0.1cm}
Ultra-high energy neutrinos ($E>10^{17}\, \rm eV$) are key particles to unveil the most powerful phenomena in the Universe. They travel without deflection nor attenuation and are a clear signature of hadronic acceleration~\cite{Ackermann_2022}. 
To target these neutrinos, new experiments with novel detection techniques are being built to reach unprecedented sensitivity, sky coverage, and angular resolution at the highest energies. Towards this goal, in-ice radio detection of neutrinos is a promising technique~\cite{Kravchenko_2003, Gorham_2021, Anker_2019}. 
When a neutrino crosses the ice, it creates a cascade of secondary particles that emit radio waves through the Askaryan emission process~\cite{Askaryan_1961}; this emission then propagates through the ice with almost no attenuation. The Askaryan Radio Array (ARA)~\cite{Allison_2016}, and the Radio Neutrino Observatory in Greenland (RNO-G)~\cite{Aguilar_2021} are two experiments with close detection concepts that rely on radio antennas buried deep inside the ice, in the South Pole or in Greenland, to target this radio emission and reconstruct the neutrino properties. Such a detection is challenging and requires identifying neutrinos solely from their radio signatures. Cosmic rays in particular are a major background since their cascades can  produce a similar radio emission in the Earth's atmosphere or directly in the ice, which can propagate and reach the deep antennas. On the other hand, the first in-situ detection of a cosmic ray would be decisive to validate  the in-ice experiments' detection principle and help calibrate the detectors. We use the innovative Monte-Carlo tool FAERIE~\cite{Kockere_2024} (Framework for the simulation of Air shower Emission
of Radio for in-Ice Experiments) to characterize the  cosmic ray radio emission as seen by deep in-ice observers. In this work, we discuss specific signatures for cosmic ray identification, based on surface antenna signals, double pulse events, and the radio signal polarization.

%We introduce our simulation setup in Section~\ref{sec:Setup}. Then in Section~\ref{sec:GeneralCharac} we show general characteristics of the cosmic ray radio signals from FAERIE simulations. Finally in Section~\ref{sec:signatures} we discuss specific signatures for cosmic ray identification, based on surface antennas signals, double pulse events and the radio signal polarization.
\vspace{-0.01cm}
\section{Antennas grid and simulation setup}~\label{sec:Setup}
\vspace{-0.1cm}
We use FAERIE to simulate the radio signal from proton showers as seen by deep in-ice observers. As the radio emission strongly depends on the primary particle energy and zenith angle, we simulate the radio emission for three energy bins $E = [10^{16.5}, 10^{17},  10^{17.5}] \rm \, eV$ and $E_{\rm max} = 10^{17.5}\, \rm eV$, and for eight zenith bins generated uniformly in $\cos{\theta}$  between $\theta  = 0^{\circ}$ and $\theta  = 50^{\circ}$, as detailed in Table~\ref{table:setup}. We consider only one azimuth angle ($\varphi = 0^{\circ}$, shower propagating towards the magnetic North) since we set the magnetic field configuration of Summit Station in Greenland ($B_x = 7.7 \,\mu {\rm T}$, $B_z =-54.1 \mu {\rm T}$), with a near-vertical direction, resulting in a mild dependency of the radio emission on the shower azimuth angle. We note that since FAERIE is computationally expensive, only one shower has been generated per set of energy and zenith bin.

\begin{table}[b]
\centering
\begin{tabular}{|c|c|}
\hline
\textbf{Energy [eV]} \rule{0pt}{2.5ex} & $10^{16.5}$; $10^{17}$; $10^{17.5}$ \\
\hline
\textbf{Zenith [°]} & 0; 20; 28; 34; 39; 43; 47; 50 \\
\hline
\textbf{Depth [m]} & 0; 40; 60; 80; 100 \\
\hline
\end{tabular}
\caption{Energy bins, zenith bins and depths of the FAERIE simulation set.}
\label{table:setup}
\end{table}

For the antenna configuration, we use squared layers of antennas at five different depths down to 100 m below the ice  surface (Table~\ref{table:setup}). Among all depths, we included surface antennas to simulate the radio emission that would be seen by the surface component of a detector such as RNO-G. The altitude above sea level is that of Summit Station with $z_{\rm ice} = 3216\, {\rm m}$. The antenna spacing is constant within a single layer but increases with the antenna depth and the shower zenith angle. The ice model follows: $n(z) = A - B\exp{-C|z|}$, where $A = 1.775$, $B =0.5019$, $C =0.03247$ for $|z| < 14.9{\, \rm m}$ and $A = 1.775$, $B =0.448023$, $C =0.02469$ for $|z| > 14.9{\, \rm m}$~\cite{RNOice}. For all our results, we filter the signal in the $[50-1000]\, \rm MHz$ range.

\begin{figure*}[tb]
\centering 
\includegraphics[width=0.49\columnwidth]{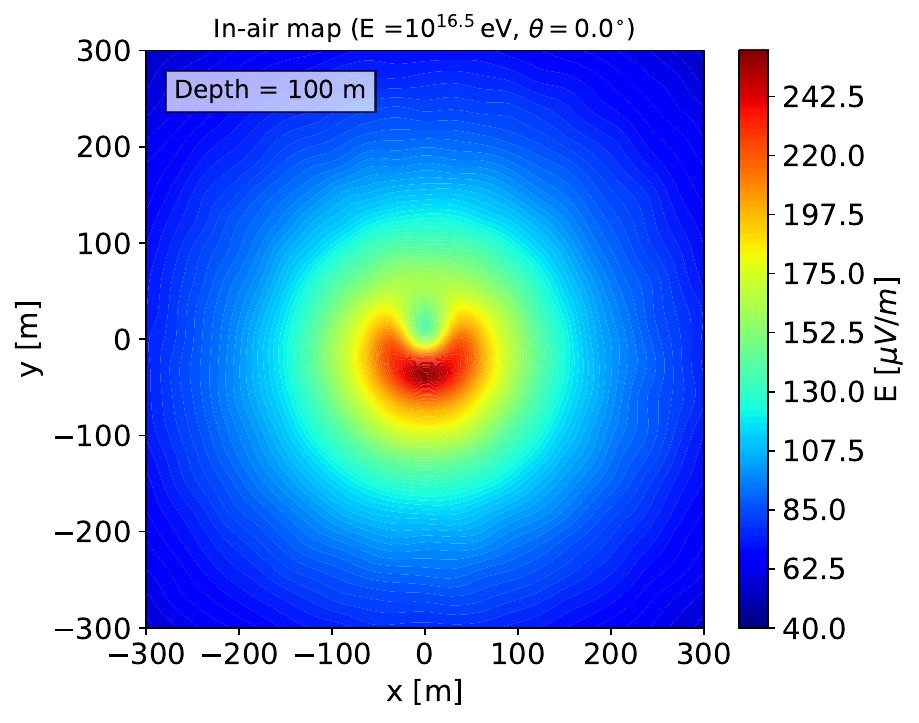}
\includegraphics[width=0.48\columnwidth]{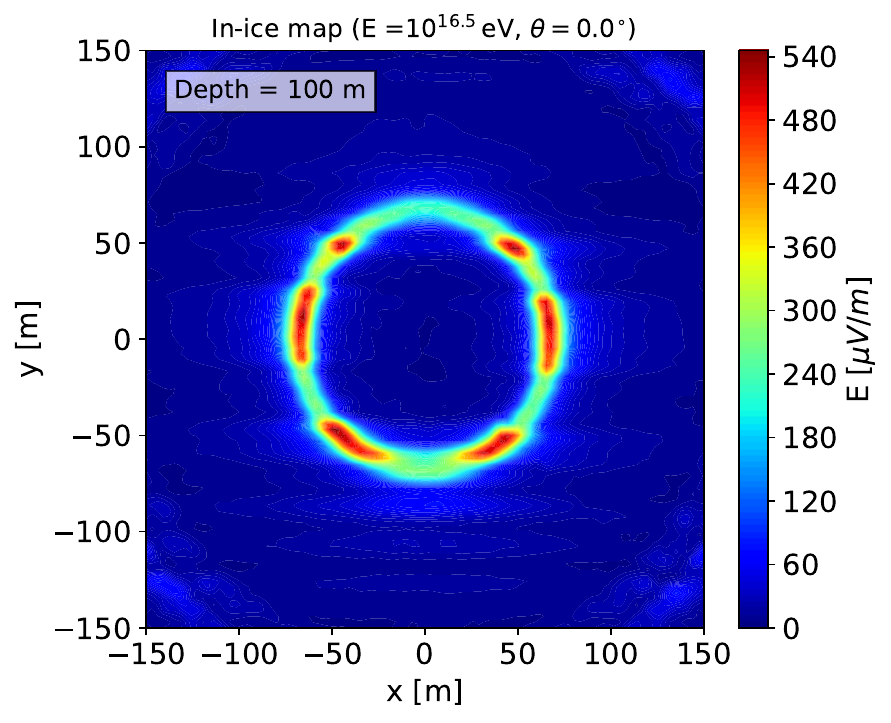}
\caption{Interpolated electric field map of ({\it left}) the in-air emission and ({\it right}) the in-ice emission seen at a depth of 100 meters below the ice surface, for a vertical proton-induced shower with primary energy $E=10^{16.5}\, \rm eV$.
}\label{fig:Footprints} \vspace{-0.3cm}
\end{figure*}

\section{General characteristics of the radio signal}~\label{sec:GeneralCharac}
\vspace{-0.3cm}

A cosmic ray interacting with the Earth's atmosphere generates a cascade of secondary particles, known as an air shower. This air shower emits coherent radio waves that can propagate to the ice and reach deep antennas. On the other hand, if the shower is energetic enough, particles  can penetrate the ice and create a secondary cascade that also emits radio waves, which can reach the deep antennas~\cite{Kockere_2024, Chiche_2024_PoS}. 

\subsection{Electric field maps}
 The FAERIE framework couples CoREAS~\cite{HuegeCOREAS} and GEANT4~\cite{ALLISON2016186} Monte-Carlo codes to simulate both the in-air and the in-ice radio emission induced by cosmic ray primaries. In Fig.~\ref{fig:Footprints}, we show the in-air (left-hand panel) and the in-ice (right-hand panel) radio emission simulated with FAERIE at a depth of 100 meters below the ice's surface for a vertical proton shower with primary energy $E = 10^{16.5}\, \rm eV$. For the in-air emission, we can observe that the emission follows a bean-shaped pattern with a stronger amplitude for the negative part of the y-axis. This feature is typically expected for the in-air emission from cosmic ray showers due to the interferences between the geomagnetic and Askaryan mechanisms~\cite{Schr_der_2017}. For the in-ice emission, we observe a  rotationally symmetric emission pattern since this signal corresponds to Askaryan emission only, which has a radial polarization. We note that for more inclined showers we mainly see an elongation of the in-air emission footprint along the azimuthal direction in which the shower propagates, while the spatial extension of the in-ice emission stays almost constant. We also mention that no significant variation of the radio signals with depth has been observed once the radio signal hits the surface, apart from a decrease of the amplitude with the distance to the emission point.

\subsection{Radiation energy}
\begin{figure*}[tb]
\centering 
\includegraphics[width=0.49\columnwidth]{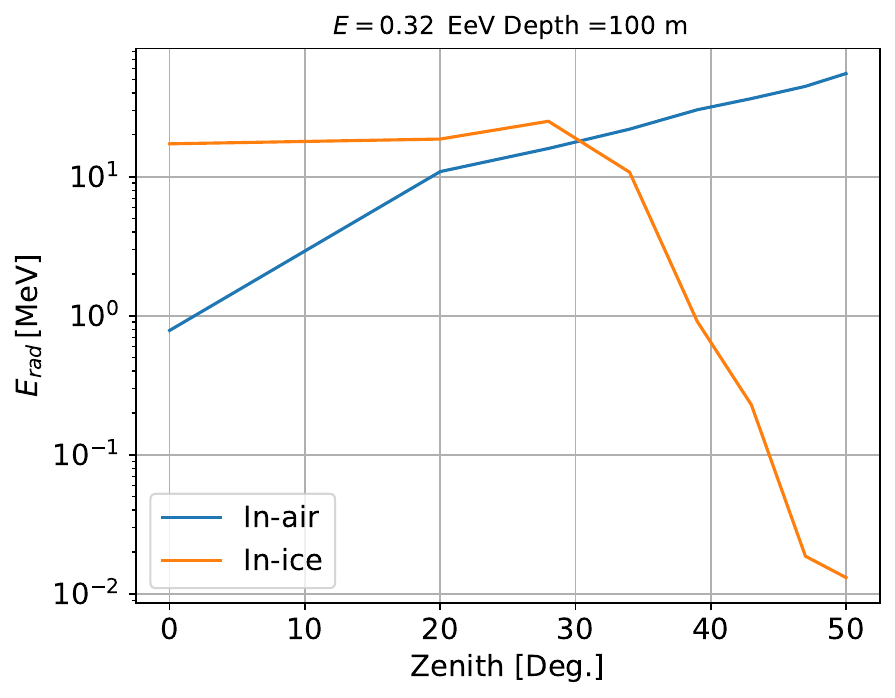}
\includegraphics[width=0.49\columnwidth]{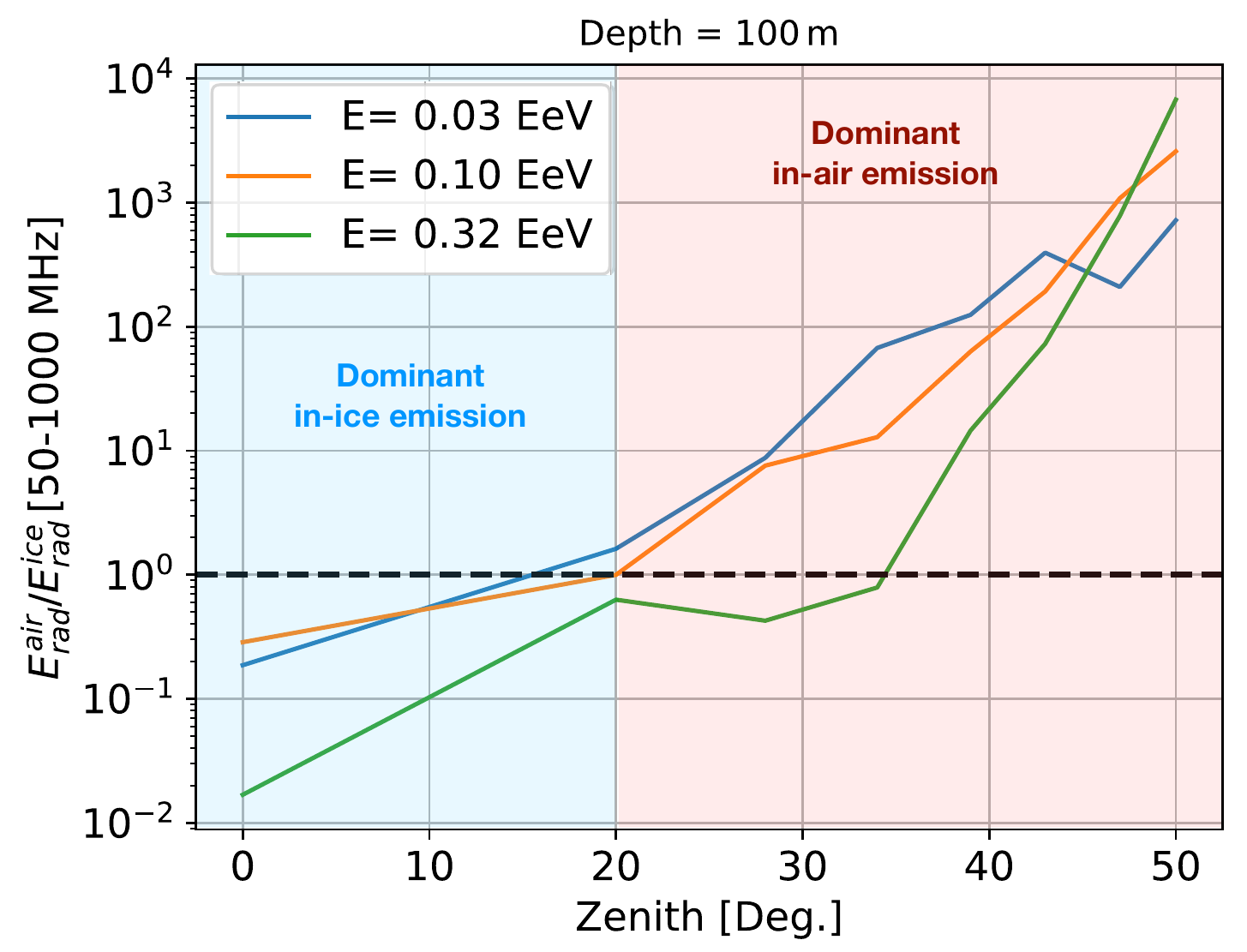}
\caption{{\it Left:} Radiation energy at a depth of 100 meters, for the in-air (blue curve) and the in-ice (orange curve) emission, as a function of the zenith angle, for proton-induced showers with primary energy $E = 10^{17.5}\, \rm eV$. {\it Right:} Ratio of the in-air and in-ice radiation energies at a depth of 100 meters, as a function of the zenith angle, for proton-induced showers with three different primary energies $[10^{16.5}, 10^{17}, 10^{17.5}]\, \rm eV$. The dashed black line indicates the limit where the ratio is equal to 1.
}\label{fig:RadiationEnergy}
\end{figure*}

The shower radiation energy is a calorimetric observable which makes it convenient to study the scaling of the radio emission with the shower parameters. It is obtained by spatially integrating the energy fluence over the antenna grid~\cite{Glaser_2016}. In the left-hand panel of Figure~\ref{fig:RadiationEnergy} we show the scaling of the in-air and in-ice radiation energies with the shower zenith angle, for proton-induced showers with a primary energy $E = 10^{17.5}\, \rm eV$. The antenna layer at $100\, \rm m$ depth was chosen to integrate the radio signal, though the results should not depend on the depth since the energy of the antenna layer is conserved through the propagation. We can observe that the in-air radiation energy increases almost continuously when increasing the zenith angle. The in-air emission is composed of geomagnetic and Askaryan emissions, with a dominant geomagnetic component. Inclined air showers develop significantly higher in the atmosphere than vertical ones and thus at lower air densities. While the energy radiated through Askaryan emission  decreases with decreasing air density,  we expect the opposite behavior for the geomagnetic emission and hence an overall increase. We also observe that the in-ice radiation energy decreases with increasing zenith angle, with a significant break around $\theta =30^{\circ}$. This is expected as more inclined showers have fewer particles that are able to reach the ice's surface with high enough energy to trigger the in-ice cascade. The break in the radiation energy is observed across all energy bins, while the zenith angle at which it occurs is shifted to lower values for lower primary energies. 
On the right-hand panel of Figure~\ref{fig:RadiationEnergy} we show the ratio between the in-air and the in-ice radiation energies. It can be seen that the in-ice component is dominant for the most vertical showers, while the in-air component becomes dominant for showers with zenith angle $\theta \gtrsim 20^{\circ}$. We also note that the transition to a dominant in-air emission is shifted to slightly higher zenith angles for higher primary energies. Indeed, the in-air radiation increases roughly linearly with the primary particle energy, while for the in-ice component, a higher primary energy allows more particles to penetrate the ice and contribute to the secondary cascade, and we expect a larger increase in the emission. We mention that these results were obtained from three single simulated showers and fluctuations of $\sim 10\%$ of the radiation energy are expected from shower-to-shower.

\section{Radio signatures for cosmic-ray identification}~\label{sec:signatures}
Identifying cosmic-ray primaries is essential to validate in-ice radio experiments' detection principle, calibrate the detectors, and perform neutrino/cosmic ray discrimination. Using FAERIE simulations, we can identify specific cosmic-ray radio signatures.

\begin{figure*}[tb]
\centering 
\includegraphics[width=0.49\columnwidth]{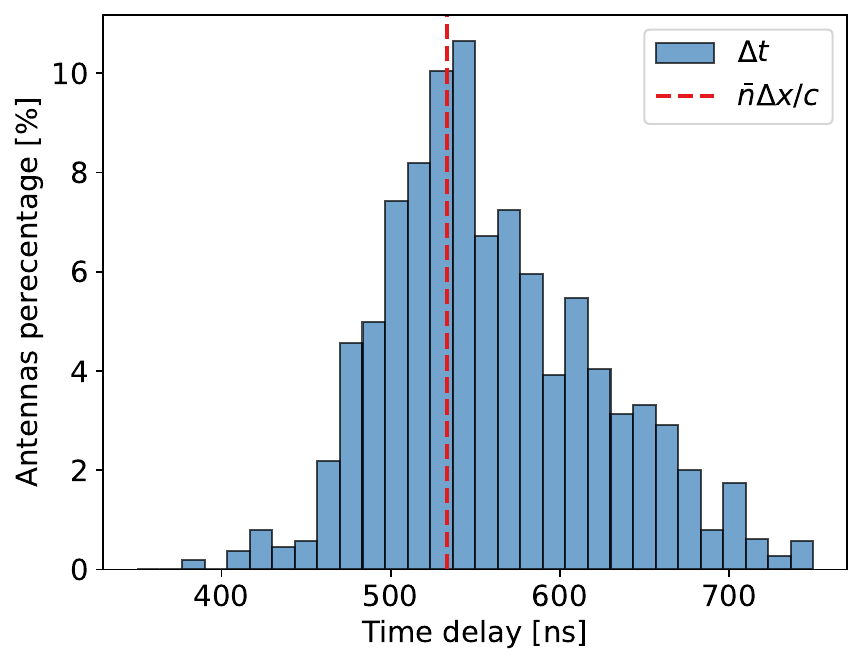}
\includegraphics[width=0.49\columnwidth]{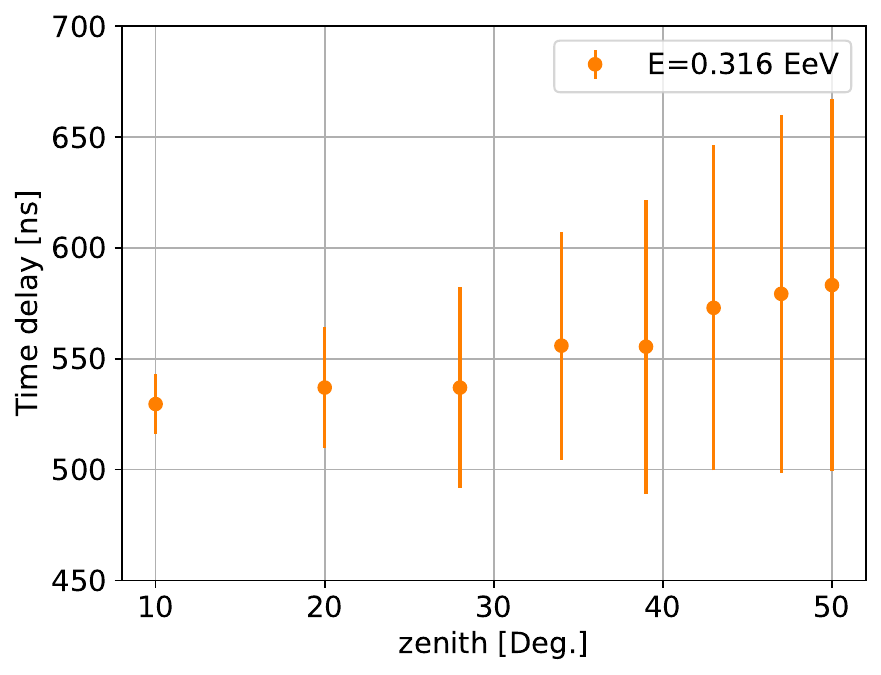}

\caption{{\it Left:} All-zenith distribution of time delays between signals from surface antennas and deep antennas at a depth of 100 meters for proton-induced showers with a primary energy $E= 10^{17.5}\, \rm eV$, from FAERIE simulations. {\it Right:} Mean (dots) and standard deviation (error bars) of the time delays between surface and 100-m deep antennas for $E=10^{17.5} \, \rm eV$.
}\label{fig:Surface}
\vspace{-0.1cm}
\end{figure*}

\subsection{Surface antenna signals}

One criteria to evaluate whether an in-ice triggered event is potentially caused by a cosmic ray–induced shower is the use of surface antennas. While neutrino primaries interact directly in the ice, cosmic rays interact in the Earth's atmosphere, which implies that their in-air emission will eventually cross the ice's surface. Experiments like RNO-G combine deep and surface antennas to identify and veto cosmic rays. Using FAERIE, we can extract the time delay between the signal detected by surface and deep antennas to evaluate in which case a correlation between the two is expected. For our estimation we: (1) set a trigger condition to use only antennas with a signal above $60 \, \mu {\rm V/m}$, (2) assume that the shower azimuth is known so that we pair deep antennas only with surface antennas that were hit before the deep one they are associated with, (3) we pair the closest deep and surface antennas that respect (1) and (2). These are, of course, simplifications to get an estimate of the time delays. In the left-hand panel of Fig.~\ref{fig:Surface}, we show the resulting time delay distribution between the in-air emission of surface and 100-m deep antennas, from FAERIE simulations, for showers with a primary energy $E = 10^{17.5}\, \rm eV$ and considering all zenith angles given in Table~\ref{table:setup}. Independently of the zenith angle we find that time delays are grouped in a narrow region with a typical extent of $0.3 \, \mu {\rm s}$. Additionally, we note that we were able to retrieve the peak of this distribution (red line) using a simple model following: $\delta t^{\rm max}=\bar{n \Delta x}/c$, where $\bar{n}=1.605$ is the average ice refractive index between 0 and a 100-m depth, $\Delta x$, the vertical distance between deep and surface antennas, and $c$ the speed of light in vacuum. The right-hand panel in Fig.~\ref{fig:Surface} shows the mean and standard deviation of the time delays as a function of the shower zenith angle. We observe that the mean time delay increases with increasing zenith angle. This is due to the fact that for more inclined showers, the radio emission arrives at the ice's surface with a more horizontal incident angle and a larger bending of the rays is expected. Another interesting trend is that the standard deviation of the time delays also increases significantly with the zenith angle. This is because inclined air showers have a more asymmetric footprint which translates into a larger dispersion of ray incident angles at the ice's surface.

\subsection{Double pulse signature}

\begin{figure*}[tb]
\centering 
\includegraphics[width=0.49\columnwidth]{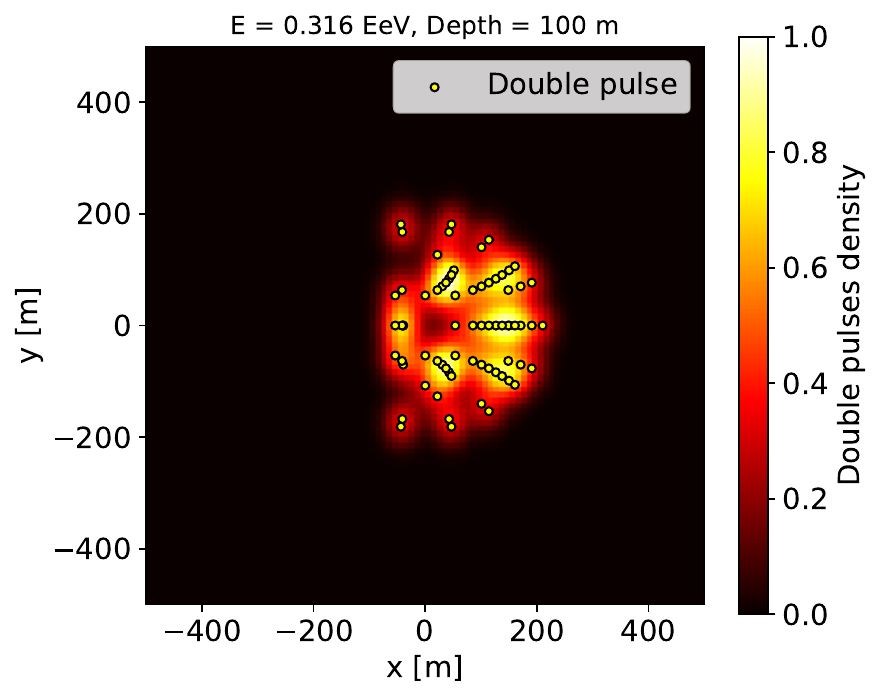}
\includegraphics[width=0.49\columnwidth]{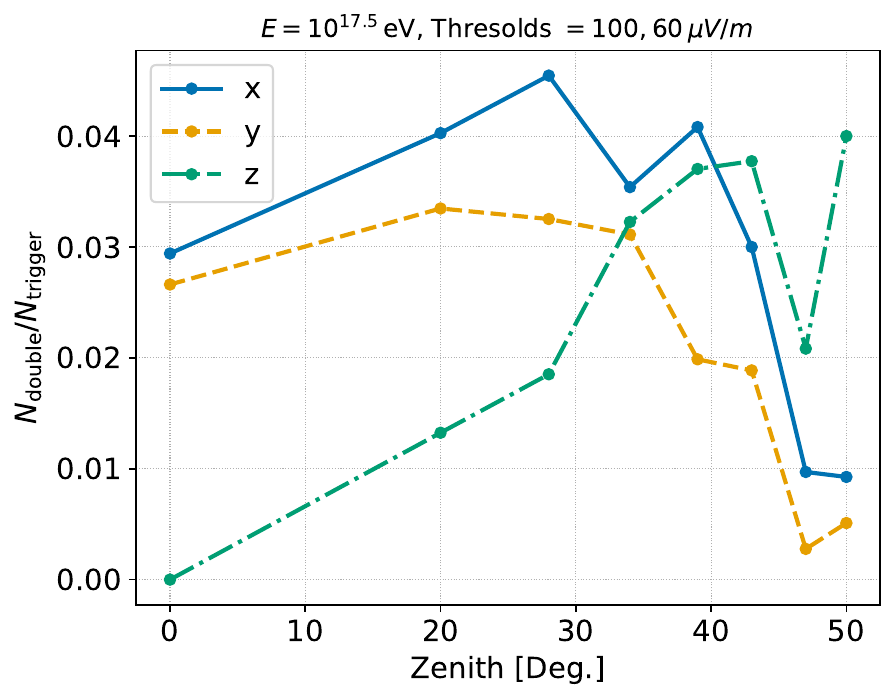}
\caption{({\it Left}) Normalized all-zenith double pulses density map at depth of 100 m for proton showers with energy $E=10^{17.5}\, \rm$ eV. ({\it Right}) Double pulses rate for each electric field channel as a function of the shower zenith angle.
}\label{fig:DoublePulses}
\end{figure*}

Another typical signature of cosmic ray-induced showers is the double pulse signal. As mentioned in Section~\ref{sec:GeneralCharac} the radio emission seen by an in-ice observer can be decomposed into two main components: the radio emission from the in-air cascade and the one from the in-ice cascade. When they overlap, these two emissions can reach the same antenna and give rise to a double pulse signature in the signal amplitude, since they usually arrive at different times. Double pulse signals are expected to be a very distinguishable signature from cosmic ray emission. Using FAERIE, we identify double pulses with a simple model, assuming that we see a double pulse  at a given antenna if: (1) there is at least one pulse with a peak amplitude $E^{\rm peak} >100 \, \mu \rm V/m$, (2) there is a second pulse with an amplitude  $E^{\rm peak} >60 \, \mu \rm V/m$. In the left-hand panel of Fig.~\ref{fig:DoublePulses}, we show a heat map of the double pulses distribution. Each yellow  dot represents a double pulse, and we can see that double pulses are concentrated in  a narrow region around the shower core ($x=0$, $y=0$). Indeed, at a depth of 100 meters, the in-ice footprint reaches its Cherenkov angle at a distance from the shower core of $\sim 80$ meters and typically does not extend beyond 200 meters which limits the spatial extent over which we see double pulses. 
We also see that the double pulse distribution is asymmetric, with more pulses along the positive x-axis. This results from two effects: vertical showers produce a symmetric in-ice emission and thus a symmetric double pulse distribution (Fig.~\ref{fig:Footprints}), while inclined showers propagating toward the positive x-axis have their negative x-side emission shadowed by ray bending, causing asymmetry. The right-hand plot shows the double pulses rate for each channel, defined as the ratio between the number of double pulses and the number of triggers. For the trigger on the individual channels, we divided the trigger thresholds on the total amplitude by a factor $\sqrt{3}$. For all channels, we observe first an increase of the double pulses rate with increasing zenith angle, followed by a decrease. As shown in the right-hand panel of Fig.~\ref{fig:RadiationEnergy}, the in-air emission is subdominant for vertical showers and therefore represents the limiting factor for the occurrence of double pulse events. However, as shown in the left-hand panel of Fig.~\ref{fig:RadiationEnergy}, the in-air radiation energy increases with zenith angle, leading to a higher likelihood of observing double pulses. Eventually, for the most inclined showers, the in-ice emission becomes subdominant, i.e., the limiting factor for double pulse rate. Since the in-ice radiation energy decreases with increasing zenith angle, the double pulses drop for the most inclined showers. Finally, we note that the differences in the relative amplitude between each channel are mainly related to the projection of the geomagnetic emission along each direction.

\subsection{Radio signal polarization}

\begin{figure*}[tb]
\centering 
\includegraphics[width=0.49\columnwidth]{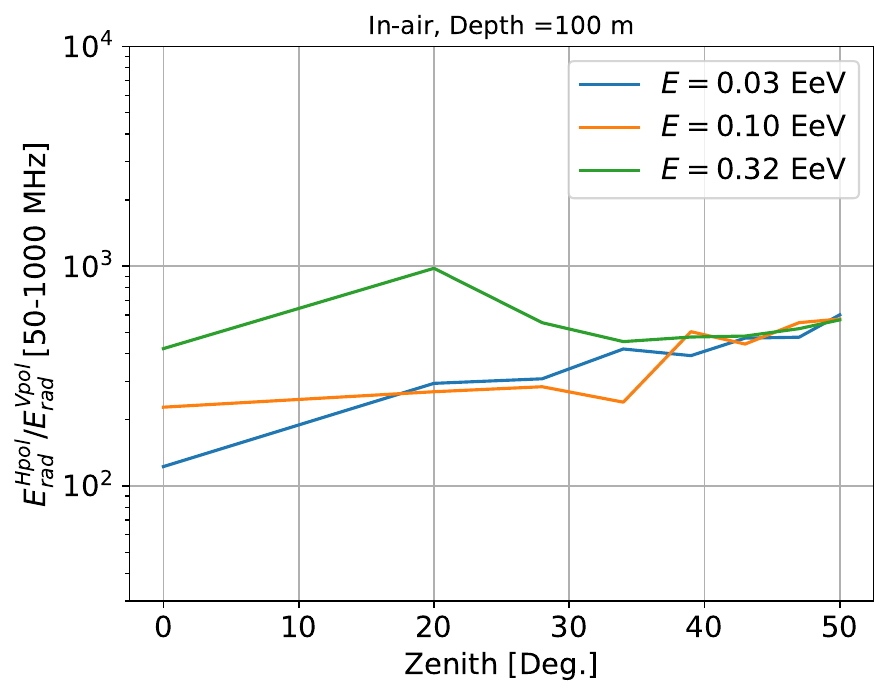}
\includegraphics[width=0.49\columnwidth]{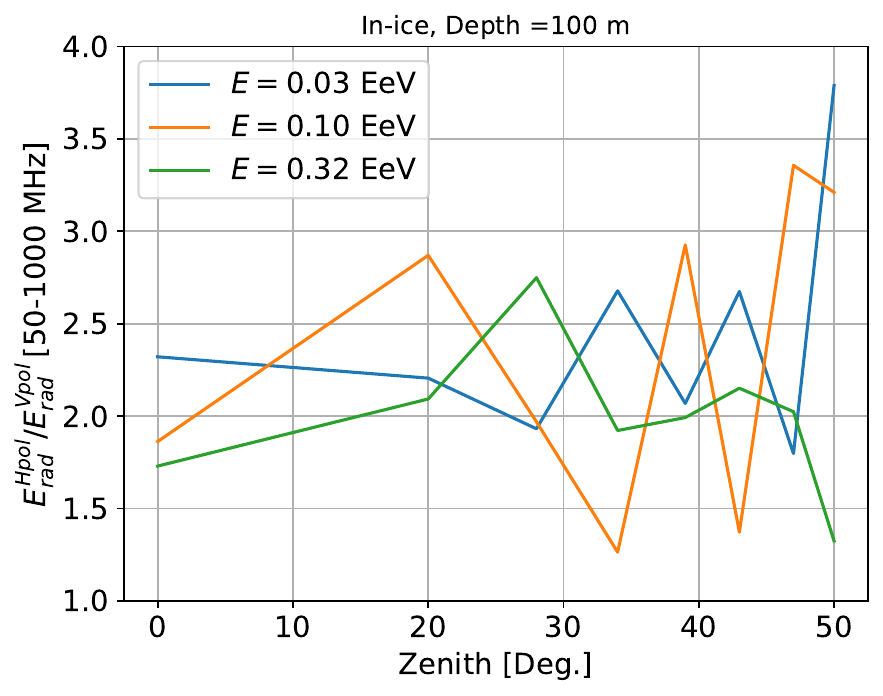}
\caption{Hpol over Vpol radiation energy ratio for ({\it left}) the in-air and ({\it right}) the in-ice emission as a function of the shower zenith angle, for different primary energies.
}\label{fig:Polarization}
\end{figure*}

Finally, we can use the  radio signal polarization to identify cosmic ray events. Typically, ARA and RNO-G use Vpol and Hpol antennas to measure the vertical and horizontal polarization of the radio signal respectively. Hence, we can use the Hpol to Vpol ratio to characterize the radio emission. Using FAERIE, we can reconstruct the radiation energy for each component of the electric field independently (x: North-South, y: West-East, z: Up-Down). Then, we estimate the radiation energy measured by the Vpol using the z-component ($E_{\rm rad}^{\rm Vpol} = E_{\rm rad}^{z}$) and the one measured by the Hpol the radiation energy along the $x$ and $y$ through: $E_{\rm rad}^{\rm Hpol} = E_{\rm rad}^{x} + E_{\rm rad}^{y}$, since Hpol antennas in RNO-G typically have a uniform azimuthal gain. In Fig.~\ref{fig:Polarization}, we show the ratio between Hpol and Vpol radiation energies for the in-air (left-hand panel) and in-ice (right-hand panel) emissions as a function of the shower zenith angle. Independently of the zenith angle, we can see that the ratio for the in-air emission is two orders of magnitude bigger than the one obtained from the in-air emission. This can be explained by the fact that the in-air emission is dominated by the geomagnetic emission which is polarized towards the $-v \times B$ direction, where $v$ is the direction of the shower axis, and $B$ is the direction of the Earth's magnetic field. For a near-vertical magnetic field, this direction happens to be mainly horizontal for the range of zenith angles we considered. On the other hand, the in-ice emission comes from the Askaryan emission which is radially polarized and therefore we expect an Hpol/Vpol ratio close to two, as seen in Fig.~\ref{fig:Polarization}. This is a strong feature, as it implies that in cases where only a single pulse is observed, the Hpol/Vpol ratio could help identify whether the signal originates from in-air or in-ice emission. This also means that in the case of double pulse signals the comparison between the Hpol/Vpol ratio of the  first and of the second pulse would be a smoking gun signature to identify cosmic rays. Eventually, this feature could help for neutrino-cosmic ray discrimination. Since neutrino emission comes from the Askaryan emission it is expected to have an Hpol/Vpol ratio  similar to the one shown in the right-hand panel of Fig.~\ref{fig:Polarization}. RNO-G expects to see neutrinos with zenith angles at least above $\theta = 40^{\circ}$ since they need to travel long enough in the ice to interact. However, as shown in Fig.~\ref{fig:RadiationEnergy}, for zenith angles above $\theta =20^{\circ}$, we expect the cosmic-ray signal to be dominated by the in-air contribution. This implies that in the range of zenith angle where neutrinos are expected we could discriminate neutrinos and cosmic rays using the radio signal polarization.

% Interesting note: surface antennas can help identify the type of emission (in-air or in-ice), for in-ice the footprint grows faster with increasing depth and the amplitude at the antenna level varies more rapidly

\vspace{-0.1cm}

\section{Conclusion}
\vspace{-0.1cm}

Using FAERIE simulations, we characterized the radio emission from cosmic ray showers seen by in-ice observers. Our results show that the in-ice component of the radio emission should be dominant for vertical showers while the in-air component becomes dominant for showers with zenith angle $\theta >20^{\circ}$. Furthermore, we evidenced that the radio signal timing, amplitude, and polarization provide promising ways to identify cosmic ray primaries through surface antennas, double pulse signals, or the Hpol/Vpol ratio. This analysis, realized at the electric field level, could then be coupled to studies that incorporate given detector responses to compare simulation results to data from ARA or RNO-G, for example. Such an implementation would be decisive to validate FAERIE simulations, identify cosmic-ray events, and validate in-ice experiments' detection principle.

%analysis could then be incorporated into studies that account for the detector response

\section*{Acknowledgments}
S. Toscano and S. Chiche are supported by the  Belgian Funds for Scientific Research (FRS-FNRS).

{\footnotesize
\begin{spacing}{0.9}

\end{spacing}}

\end{document}